\documentclass[aps, prl, twocolumn, floatfix, superscriptaddress, reprint, 10pt]{revtex4-1}  
\usepackage{graphicx}
\usepackage[usenames]{color}
\usepackage{amsmath}
\usepackage{amssymb}

\usepackage{bm}
\usepackage{physics}
\usepackage{verbatim}
\usepackage{graphics}

\newcommand{\xj}{\mathcal{X}_{j}}
\newcommand{\xk}{\mathcal{X}_{k}}
\newcommand{\br}{\mathbf{r}}

\newcommand{\dbr}{\textrm{d}\br}

\newcommand{\drhat}{\textrm{d}\hat{R}}

\newcommand{\CX}{\mathcal{X}}

\newcommand{\CA}{\mathcal{A}}
\newcommand{\CB}{\mathcal{B}}

\begin{document}
\title{Feature Optimization for Atomistic Machine Learning Yields A Data-Driven
Construction of the Periodic Table of the Elements}
\author{Michael J.~Willatt}
\thanks{These authors contributed equally to this work. Corresponding author:
michael.willatt@epfl.ch}
\affiliation{Laboratory of Computational Science and Modeling, IMX, \'Ecole Polytechnique F\'ed\'erale de Lausanne, 1015 Lausanne, Switzerland}
\author{F\'elix Musil}
\thanks{These authors contributed equally to this work. Corresponding author:
michael.willatt@epfl.ch}
\affiliation{Laboratory of Computational Science and Modeling, IMX, \'Ecole Polytechnique F\'ed\'erale de Lausanne, 1015 Lausanne, Switzerland}
\affiliation{National Center
for Computational Design and Discovery of Novel Materials (MARVEL)}
\author{Michele Ceriotti}
\affiliation{Laboratory of Computational Science and Modeling, IMX, \'Ecole Polytechnique F\'ed\'erale de Lausanne, 1015 Lausanne, Switzerland}

\begin{abstract}
Machine-learning of atomic-scale properties amounts to extracting correlations between
structure, composition and the quantity that one wants to predict.  Representing the input
structure in a way that best reflects such correlations makes it possible to improve the
accuracy of the model for a given amount of reference data. When using a description of
the structures that is transparent and well-principled, optimizing the representation
might reveal insights into the chemistry of the data set.  Here we show how one can
generalize the SOAP kernel to introduce a distance-dependent weight that accounts for the
multi-scale nature of the interactions, and a description of correlations between chemical
species.  We show that this improves substantially the performance of ML models of
molecular and materials stability, while making it easier to work with complex,
multi-component systems and to extend SOAP to coarse-grained intermolecular potentials.
The element correlations that give the best performing model show striking similarities
with the conventional periodic table of the elements, providing an inspiring example of
how machine learning can rediscover, and generalize, intuitive concepts that constitute
the foundations of chemistry. \textit{The following article appeared in Physical Chemistry
Chemical Physics, 20, 29661 (2018) and may be found at
http://dx.doi.org/10.1039/C8CP05921G}
\end{abstract}
\maketitle

In the last few years, statistical regression techniques have gained an important place in the toolbox of atomic-scale modelling, making it possible to approximate effectively the properties of systems computed with accurate but demanding electronic structure methods based on a small number of reference calculations~\cite{behl-parr07prl,bart+10prl,rupp+12prl}. 
It is fair to say that most of the recent progress in this field has been associated with the development of representations that encode the fundamental symmetries of the system~\cite{bart+13prb, behl-parr07prl,glie+17prb,Grisafi2018,Glielmo2018, VonLilienfeld2018}. 
After symmetries have been accounted for, however, there is still considerable freedom in how to define the details of an atomic-scale representation.
Optimizing the input representation can improve substantially the performance of the regression, by adapting it to the specific structure-property relations associated with a given problem. 
What is more, in the process one can often recognize correlations that rely on intuitive information on such structure-property relations. 
In this paper we consider the smooth overlap of atomic positions (SOAP) representation -- a popular representation of atomic structure that has been successfully used to build interatomic potentials~\cite{bart+13prb2,Deringer2017,Dragoni2018}, predict molecular properties~\cite{Bartok2017} and visualize structural motifs~\cite{De2016,de+16jci,musi+18cs} -- and extend it by adapting the representation to the intrinsic length scales of atomic interactions, and by considering ``alchemical'' correlations between chemical species, which make it possible for instance to exploit the similar behavior of different elements to accelerate learning in very chemically heterogeneous data sets.
Not only do these extensions improve significantly the performance of SOAP representations, but they do indeed offer insights into the chemistry of the system, for instance providing a data-driven representation of the similarity between elements that is reminiscent of the periodic table of the elements.

\section{Methods}

Many machine learning schemes have been used to link structures and properties~\cite{VonLilienfeld2013, fabe+17jctc}, including more or less sophisticated neural networks~\cite{bhol+07nimpr, behl11jcp, Chmiela2017, smit+17cs, Zhang2018}. 
Based on the few comparative studies that have appeared in the literature~\cite{fabe+17jctc,nguy+18jcp,Qu2018}, it appears that, when it comes to predicting atomic-scale properties, simple regression techniques such as kernel ridge (Gaussian process) regression perform as well as or better than their more sophisticated counterparts.
Given our focus on structure representations, in this work we used kernel ridge regression (KRR), in which the properties of a structure $\CA$ are written as a linear combination of non-linear kernel functions $K(\CX,\CX')$ that evaluate the similarity between two structures, i.e.
\begin{equation}
y(\CA) = \sum_M x_M K(\CA,\CX_M),
\end{equation}
where $\CX_M$ correspond to a set of reference atomic structures, and $x_M$ are weights that can be optimized by minimizing the discrepancy between the predictons $y(\CA)$ and the actual values computed on a set of training structures.
The details of KRR have been discussed at length elsewhere~\cite{rasm06book, Bishop2016} and so here we will focus instead on the definition and optimization of the kernel function. 

\subsection{SOAP in a bra-ket notation}

The representer theorem guarantees that every well-behaved kernel corresponds to a scalar product between feature vectors that associate each input to a point in a -- possibly infinite dimensional -- Hilbert space, $K(\CX,\CX')=\boldsymbol{\Phi}(\CX)^T\boldsymbol{\Phi}(\CX')$~\cite{Cuturi2010}. 
The Dirac notation provides a convenient formalism to express vectors in a Hilbert space in an abstract way that is basis-set independent. This makes it very suitable to express results in quantum mechanics, and -- due to the similar algebraic structure -- in the context of machine-learning based on kernel ridge regression~\cite{rasm06book, arxiv_p2}, where one can write $K(\CX,\CX')=\bra{\CX}\ket{\CX'}$. 

In the SOAP framework~\cite{Bartok2013}, spherical environments centered on each atom in the system are expressed as densities, which are constructed by superimposing Gaussians functions $g(\br)$, centered on each atom.  We write such an atom density as a position representation of a ket $\ket{\xj}$,
\begin{equation}
    \bra{\br}\ket{\xj} = \sum_i f_c(r_{ij})
    g(\br-\br_{ij})\ket{\alpha_i},
    \label{eq:dirac-rxj}
\end{equation}
where $\br_{ij}$ is the displacement vector between atoms $i$ and $j$ and we have introduced orthonormal element kets $\ket{\alpha}$ that represent the chemical nature of atoms, and a smooth cutoff function $f_c(r)$ that limits the density to a neighborhood of atom $j$.
We can collect together the density from all the atoms of the same species to define an element-specific density 
\begin{equation}
 \bra{\alpha \br}\ket{\xj} \equiv   \psi^\alpha_{\xj}(\br) = \sum_{i \in \alpha} f_c(\br_{ij})
    g(\br-\br_{ij}),
    \label{eq:dirac5}
\end{equation}
where we used the fact that element kets are taken to be orthogonal.
This makes it possible to write the position representation of $\ket{\xj}$ as a sum over the elements
\begin{equation}
    \bra{\br}\ket{\xj} = \sum_\alpha \psi^\alpha_{\xj}(\br)\ket{\alpha}.
    \label{eq:dirac4}
\end{equation}

In the original formulation of SOAP~\cite{Bartok2013}, the atom density is expressed by expanding the environmental density in a basis of orthogonal radial basis functions $R_{n}(r)$ and spherical harmonics $Y_{m}^{l}(\hat{\br})$, 
\begin{equation}
\bra{\alpha nlm}\ket{\xj} = \int \dbr \, R_{n}(r)Y_{m}^{l}(\hat{\br})\bra{\alpha\br}\ket{\xj}.
    \label{eq:anlm-xj}
\end{equation}
This representation is invariant to permutations of atoms of the same kind, and to rigid translations. It is not, however, rotationally invariant, and so the kernel built as the overlap between two environments would not be consistent with one fundamental physical symmetry.
To remedy this shortcoming, one can average the kernel over the $SO(3)$ rotation group to obtain
\begin{equation}
K^{(\nu)}(\xj,\xk) = \int \drhat \bra{\xj}\hat{R}\ket{\xk}^\nu. \label{eq:soap-r-int}
\end{equation}
A remarkable result of the SOAP framework is that the representations that are associated with this kernel can be computed explicitly.
For the case with $\nu=2$, the SOAP representations correspond to the power spectrum,
\begin{equation}
    \bra{\alpha n\alpha' n'l}\ket{\xj^{(2)}} \propto \frac{1}{\sqrt{2l+1}} \sum_{m} \bra{\alpha nlm}\ket{\xj}^{\star}
    \bra{\alpha' n'lm}\ket{\xj},
\end{equation}
where $^{\star}$ denotes complex conjugation. 
One can show that 
\begin{equation}
 \bra{\xj^{(2)}}\ket{\xk^{(2)}}
=\sum_{\alpha n\alpha' n'l}\bra{\xj^{(2)}}\ket{\alpha n\alpha' n'l}\bra{\alpha n\alpha' n'l}\ket{\xk^{(2)}} 
\end{equation}
is precisely the rotationally-averaged kernel \eqref{eq:soap-r-int} for $\nu=2$, and that it captures the 3-body correlations between atoms within the environment~\cite{Glielmo2018}.

To complete our summary of the SOAP framework, we should mention that in many applications thus far the $SO(3)$ vectors have been normalised,
\begin{equation}
    \ket{\xj^{(\nu)}}/\sqrt{\bra{\xj^{(\nu)}}\ket{\xj^{(\nu)}}} \to \ket{\CX_{j}^{(\nu)}},
\end{equation}
and raised to an integer power $\zeta$, 
\begin{equation}
    \underbrace{\ket{\xj^{(\nu)}}\otimes \ket{\xj^{(\nu)}}\otimes \dots\otimes \ket{\xj^{(\nu)}}}_{{\zeta}} \to \ket{\xj^{(\nu)}},
    \label{eq:soap-zeta}
\end{equation}
which makes it possible to go beyond the body order implied by $\nu$ in the definition of the environmental kernel, avoiding the complications of higher-order SOAP representations.

Having constructed the $SO(3)$ vectors, there are a variety of ways to obtain a global correlation measure between atomic configurations~\cite{De2016}. The simplest approach (which we follow in this work, and is appropriate to learn properties that can be decomposed in atom-centered contributions) is the average kernel
\begin{equation}
    K^{(\nu)}(\CA,\CB) = \frac{1}{N_{\CA}N_{\CB}} \sum_{j\in A}\sum_{k\in B}
    \bra{\CX_{j}^{(\nu)}}\ket{\CX_{k}^{(\nu)}},
    \label{eq:global1}
\end{equation}
where $N_{\CA}$ ($N_{\CB}$) is the number of environments in $\CA$ ($\CB$).

\section{Generalising the SOAP environmental kernel}

The SOAP formalism provides an elegant framework to construct a rotationally-invariant representation of the atomic density that can be used for machine-learning purposes.   While the formalism provides a complete representation of structural correlations of a given order within an atomic environment, the quality and the computational cost of the regression scheme can be improved substantially in practice by modifying the representation so that it incorporates some degree of chemical intuition.  For instance, the combination of multiple kernels corresponding to different interatomic distances has been shown to improve the quality of ML models\cite{Bartok2017}, and the use of an alchemical kernel matrix to describe the similarity of different elements has been shown to be beneficial as well~\cite{De2016,Bartok2017}. 

\subsection{Radially-scaled kernels}
\label{sub:radial}

In a system with relatively uniform atom density, the overlap between environments $\bra{\xj}\ket{\xk}$ is dominated by the region farthest from the centre.  This could be regarded as rather unphysical, since the interactions between atoms decay with distance, and so the closest atoms should give the most significant contribution to the properties, and is reflected in the observation that multi-scale kernels tend to perform best when very low weights are assigned to the long-range kernels~\cite{Bartok2017,paruzzo2018chemical}. 
Likewise, a scaling of the weights of different atomic distances within an environment has been shown to be beneficial when using ML to predict atomic-scale properties using a different density-based representation~\cite{Faber2018}.

One could modify SOAP features to compensate for this effect by multiplying the atomic probability amplitude~\eqref{eq:dirac-rxj} with a radial scaling $u(r)$.\cite{Huang2016}
For ease of implementation, however, we apply the scaling directly in the definition of $\psi_{\xj}(\br)$, using $u(r)$ to determine weights associated with each atom in the environment,
\begin{equation}
\bra{\alpha\br}\ket{\xj} = \sum_{i\in \alpha} f_c(r_{ij}) u (r_{ij})
    g(\br-\br_{ij}).    \label{eq:dirac-rxj-ur}
\end{equation}
While this construction is an accurate realisation of a density scaling only when the width of the atomic Gaussians is small compared to the variation of $u(r)$, it provides a simple way to test the general idea that requires minimal changes to existing SOAP code.\cite{QUIP}
One should also consider that the atom that sits at the centre of the environment has a special status in the SOAP framework. While atoms in the environment provide information on the structural correlations, the $j$-th atom sits at the centre of the environment by construction. 
As a consequence, it is best to treat separately the weight $u_0$ associated with the central atom, i.e. to consider
\begin{equation}
\bra{\br}\ket{\xj} = u_0 g(\br) \ket{\alpha_j}+ \sum_{i \ne j} f_c(r_{ij}) u (r_{ij})
    g(\br-\br_{ij}) \ket{\alpha_i}.    \label{eq:dirac-rxj-ur-c0}
\end{equation}

\subsection{Alchemical kernels}

In the presence of multiple elements, the Dirac notation makes it evident that SOAP representations consider each element separately, and do not include a notion of similarity between different elements. This makes the computational cost grow steeply with chemical diversity, and makes it impossible to exploit the similar behavior of different elements across the periodic table. In Refs.~\cite{De2016,Bartok2017} it was shown that extending SOAP with an alchemical kernel $\kappa_{\alpha\alpha'}$ coupling different elements improved the learning efficiency. It however led to a large increase of the computational cost, as it required considering more terms in the scalar product between two representations,
\begin{equation}
\begin{split}
\bra{\xj^{(2)}}\ket{\xk^{(2)}}_\kappa = & 
\sum_{\alpha \beta n\alpha' \beta' n' l}
\bra{\xj^{(2)}}\ket{\alpha n \alpha' n' l} \\
\times & 
\kappa_{\alpha\beta} \kappa_{\alpha'\beta'} \bra{\beta n \beta' n' l}\ket{\xk^{(2)}}.
\end{split}
\label{eq:alchemy-kappa}
\end{equation}
The bra-ket notation suggests that $\kappa_{\alpha\alpha'}$ serves essentially the purpose of an operator coupling the elements $\ket{\alpha}$ and $\ket{\alpha'}$. 

In this spirit, one can write a decomposition of $\kappa$,
\begin{equation}
\kappa_{\alpha\alpha'} = \sum_{J=1}^{d_J} u_{\alpha J } u_{\alpha' J},
\end{equation}
where the coefficients can be seen as the components of the element kets on an ``elemental feature'' $\ket{J}$, i.e. $u_{\alpha J}=\bra{J}\ket{\alpha}$.
One can then rewrite Eq.~\eqref{eq:alchemy-kappa} as 
\begin{equation}
    \label{eq:alchemy-kernel}
\bra{\xj^{(2)}}\ket{\xk^{(2)}}_\kappa =
    \sum_{JnJ'n'l} \bra{\xj^{(2)}}\ket{JnJ'n'l} \bra{JnJ'n'l}\ket{\xk^{(2)}},
\end{equation}
in which we have introduced  a partially contracted version of the original representation,
\begin{equation}
    \begin{split}
    \nonumber
    \bra{JnJ'n'l}\ket{\xj^{(2)}} =& \sum_{\alpha \alpha'} u_{J\alpha}u_{J'\alpha'} \bra{\alpha n \alpha' n'l}\ket{\xj^{(2)}}.
    \label{eq:alchemy-feat}
    \end{split}
\end{equation}

The transformed SO(3) vector components can be written in terms of the components of $\ket{J}$ in the element basis, $u_{J\alpha} = \bra{J}\ket{\alpha}$. If the number of basis kets $d_J$ is smaller than the number of elements under consideration, then the effective SO(3) vectors occupy a smaller space than $\{\ket{\alpha n\alpha' n'l}\}$.   
This low-dimensionality representation of the chemical space can help improve the accuracy of a ML model in the presence of a large number of elements, and can also translate into substantial savings in terms of memory usage and computational effort. It does, however, break the sparsity of the representations, which can negatively affect the computational efficiency for some systems.

Note that this transformation can also be expressed directly in terms of the atom density, i.e. one can write
\begin{equation}
\bra{J\br}\ket{\xj} = \sum_\alpha u_{J\alpha} \psi_{\xj}^\alpha(\br).
\end{equation}
In the case with $d_J=1$ this formulation is analogous to several recent attempts to mitigate the complexity of ML models including many chemical species~\cite{nong+17prb,gast+18jcp,Huo2017} by representing heterogeneous systems with a single density, and different weights assigned to various elements. 
Rewriting the SOAP environmental kernel as Eq.~(\ref{eq:alchemy-kernel}) makes it possible to consider the  $u_{J\alpha} = \bra{J}\ket{\alpha}$ as  optimizable parameters, to improve the performance of the alchemical kernel $\kappa_{\alpha\alpha'} = \sum_{J} u_{\alpha J}u_{J \alpha'}$, or to force it to be low-rank.
Different strategies can be used to determine the optimal $u_{J\alpha}$. Here we propose a cross-validation scheme that exploits the scalar-product nature of the SOAP kernel to re-cast one part of the problem as linear regression,\cite{Bishop2016} which we discuss in detail in the Supporting Information. 

\subsection{Multiple-kernel learning}

We have shown that by manipulating the form of the SOAP kernel, e.g. by including a radial scaling, by introducing correlations between elements, or by adjusting other hyperparameters, such as the cutoff radius or the shape of the atomic Gaussian functions, it is possible to obtain different perspectives of the structural correlations, and to tune them to give the best possible performance in a regression task.
Determining the most effective representation of a given input is typically what deep neural networks are thought to excel at~\cite{goodfellow2016deep}, and exploring this possibility will be the subject of future investigation. 
Remaining in the context of kernel ridge regression, one can attempt a different approach to further improve the performance of the regression. As done in Ref.~\cite{Bartok2017}, one can build a composite kernel out of a selection of different models, i.e.
\begin{equation}
K(\CA,\CB) = \sum_\aleph w_\aleph K_\aleph(\CA,\CB).
\label{eq:multi-k}
\end{equation}
This multiple-kernel model makes it possible to find the best combination of different representations of the atomic environments, using short and long-range, 2 and 3-body, radially-scaled and alchemically-contracted terms.
In a Gaussian Process Regression language, each model is meant to contribute $\sqrt{w_\aleph}$ to the variance of the target property. The weights can be set manually based on an intuitive understanding of how they contribute to a property, or -- more simply -- optimized by cross-validation. 
Note that such combined kernels can still be seen as an explicit inner product between representations. In other words, taking sums of multiple kernels can be interpreted equivalently as generalisations of kernels, or as generalisations of representations that take the form
\begin{align}
    \ket{\CX} = \sqrt{w}_{1} \ket{\CX^{1}} \oplus \sqrt{w}_{2} \ket{\CX^{2}} \oplus \dots,
\end{align}
where $\oplus$ denotes concatenation. 

\section{Results and discussion}

Having discussed different ways SOAP representations can be modified to represent in a more efficient way structure-property relations in complex data sets, we now verify what the practical implications of such modifications are.
In order put these ideas to the test, we chose two data sets, one of which contains geometrically diverse, isolated organic molecules while the other contains elementally diverse inorganic crystals.

\noindent{\bf The QM9 data set} is a collection of about 134k DFT-optimized (B3LYP/6-31G) structures of small organic molecules.\cite{Ramakrishnan2014} Ea ch molecule contains up to nine heavy atoms (C, N, O and F) in addition to H. While the data set comprises only five atomic elements, it encompasses 621 distinct stoichiometries and is therefore very diverse geometrically.  We followed Ref.~\citenum{Ramakrishnan2014} by removing all the 3,054 structures that failed the SMILES consistency test. 
The QM9 data set has been used in many pioneering studies of machine learning for molecules, notably for the demonstration of the predictive power of methods based on Coulomb matrices\cite{Ramakrishnan2015a, Huang2016}, radial distribution functions\cite{fabe+18jcp} and SOAP.\cite{Bartok2017} It has also been used together with deep-learning schemes, such as Sch-Net~\cite{schu+18jcp} and HIP-NN~\cite{lubb+18jcp}. QM9 is a very heterogeneous data set, with some stoichiometries being heavily represented, and some considerably less sampled (e.g. F-containing compounds). This, together with the fact that it has been thoroughly benchmarked with several different  representations and regression strategies~\cite{fabe+17jctc}, makes it an ideal benchmark to demonstrate the improved learning that is made possible by the scheme we introduce here.

\noindent{\bf The elpasolite data set} comprises about 11k DFT-optimized quaternary structures with stoichiometry ABC\textsubscript{2}D\textsubscript{6} (elpasolite AlNaK\textsubscript{2}F\textsubscript{6} being the archetype). We have used the elpasolite data set of Faber \textit{et al.}\cite{Faber2016} in which the four elements constituting each structure were chosen from the 39 main group elements H to Bi.  The DFT-relaxed geometries of each structure in the elpasolite crystals are almost identical which means that the data set is geometrically uniform but elementally diverse.

\subsection{Training data selection}

For each data set, we randomly selected two subsets: an optimization set (A) to be used to determine the hyperparameters of the model by cross-validation, and the other (B) to be used for training and testing. All of the optimizations discussed in this article (radial scaling, alchemical kernel learning and multiple-kernel learning) were performed on the A set. Once each optimization was performed, we randomly shuffled and partitioned set B multiple times to produce training set and test set pairs. In order to account for the variability of the model accuracy with respect to the composition of the training and test sets, we averaged over the learning curves for each pair to create the figures presented here.

\begin{figure}[tbp]
\begin{center}
    \includegraphics[scale=0.50]{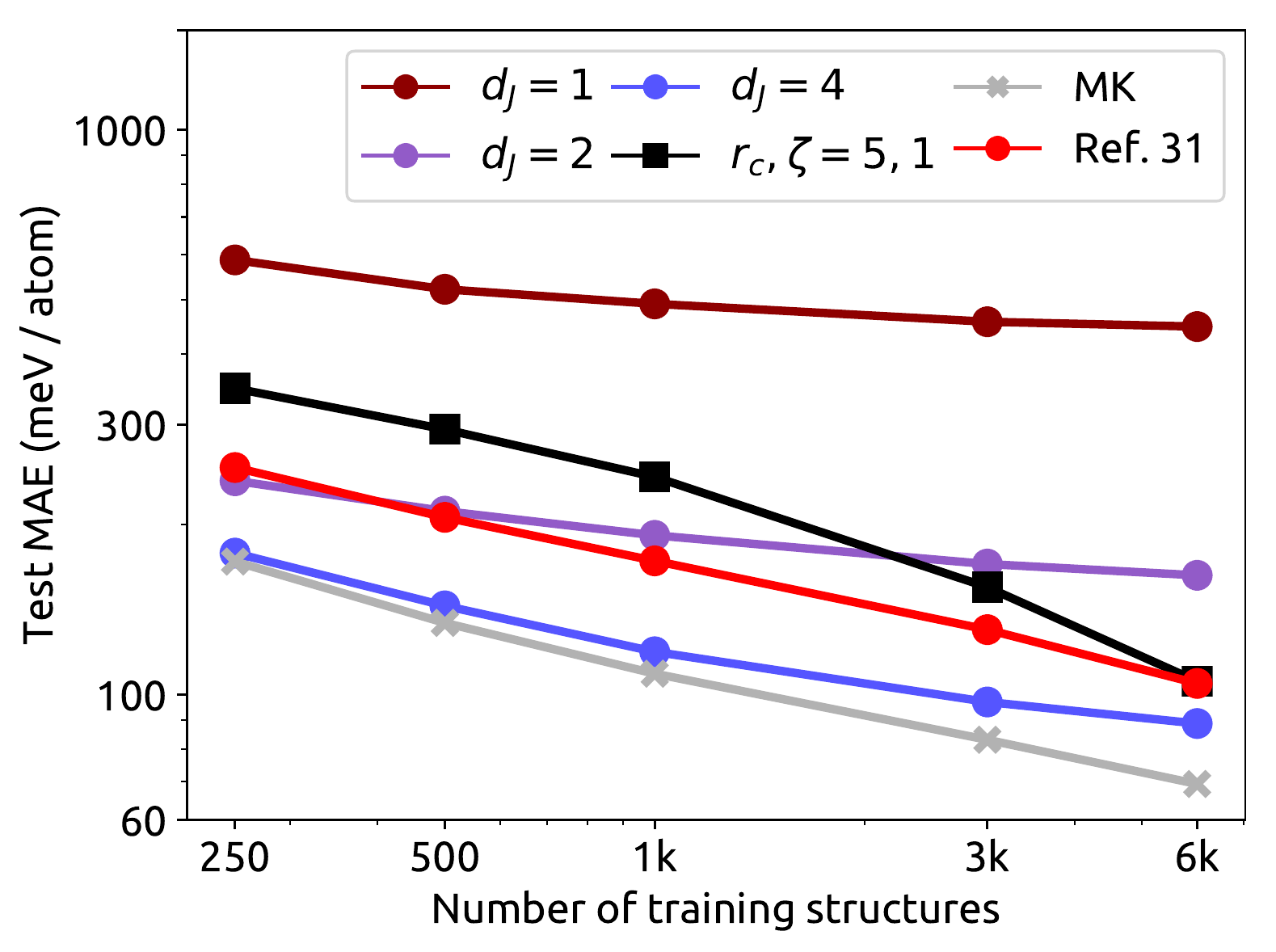}
\end{center}
\caption{Learning curves for the elpasolite crystals. The standard SOAP curve is shown in black, the best curve from Ref.~\cite{Faber2018} is shown in bright red and the optimized curves are shown in dark red ($d_J=1$), purple ($d_J=2$) and blue ($d_J=4$). For each of these models, the kernels were constructed with $r_{c}=5$\AA \, and $\zeta=1$. The multiple-kernel model (shown in grey) combines three standard SOAP kernels ($\zeta=1$, $r_{c}=4$; $\zeta=1$, $r_{c}=6$; $\zeta=4$, $r_{c}=6$) and one optimized kernel ($d_{J}=4$, $\zeta=1$, $r_{c}=5$) in the ratio $4:3:1:220$. All of the kernels were constructed with $\nu=2$, $n_\text{max}=12$ radial basis functions and $l_\text{max}=9$ non-degenerate spherical harmonics. Error bars are omitted because they are as small as the data point markers.
}
\label{fig:elpasolites-lc}
\end{figure}

\subsection{Reduced-dimensionality alchemical kernels}

For the elpasolite crystals, our optimization set contained 2k structures and the remainder were used to construct five training and test set pairs at random (6k and 2k structures respectively). Figure \ref{fig:elpasolites-lc} shows the averaged learning curves. The reference curve (bright red line) was taken from Ref.~\cite{Faber2018} and corresponds to recently-proposed density-based representations.
The dark red, purple and blue curves show the result of optimizing the alchemical kernel, which we did by initializing low-dimensional $u_{J\alpha}$ based on the $d_J$ principal components of the  alchemical kernel,
\begin{equation}
\label{eq:kappa-er}
\kappa_{\alpha\beta}=e^{-(\epsilon_\alpha-\epsilon_\beta)^2/2\sigma_\epsilon^2 - (r_\alpha-r_\beta)^2/2\sigma_r^2 },
\end{equation}
where $\epsilon_\alpha$ and $r_\alpha$ correspond to Pauling atomic electronegativity and van der Waals radius for the element $\alpha$. 
The values of $u_{J\alpha}$ were then optimized with an iterative scheme working in the primal formulation of ridge regression for $\zeta=1$ (see SI). 

Reducing the dimensionality of the SOAP representations by three orders of magnitude with $d_{J}=1$  leads to a poor learning rate (dark red line). The learning behaviour is much improved with $d_J = 2$ (purple line), which corresponds to a reduction in the dimensionality of the SOAP representations by a factor of 380. For fewer than 2k structures, the performance is better than standard SOAP (black line), but the learning rate gradually decreases (saturation) as the number of training structures increases. This suggests that the $d_{J}=2$ representation is unable to represent diversity adequately in large sets of structures because of its low dimensionality, in much the same way as reducing $\zeta$ has been found to lead to saturation in SOAP models trained on the QM9 data set.\cite{Bartok2013a}

\begin{figure}[tbp]
\begin{center}
    \includegraphics[width=1.0\linewidth]{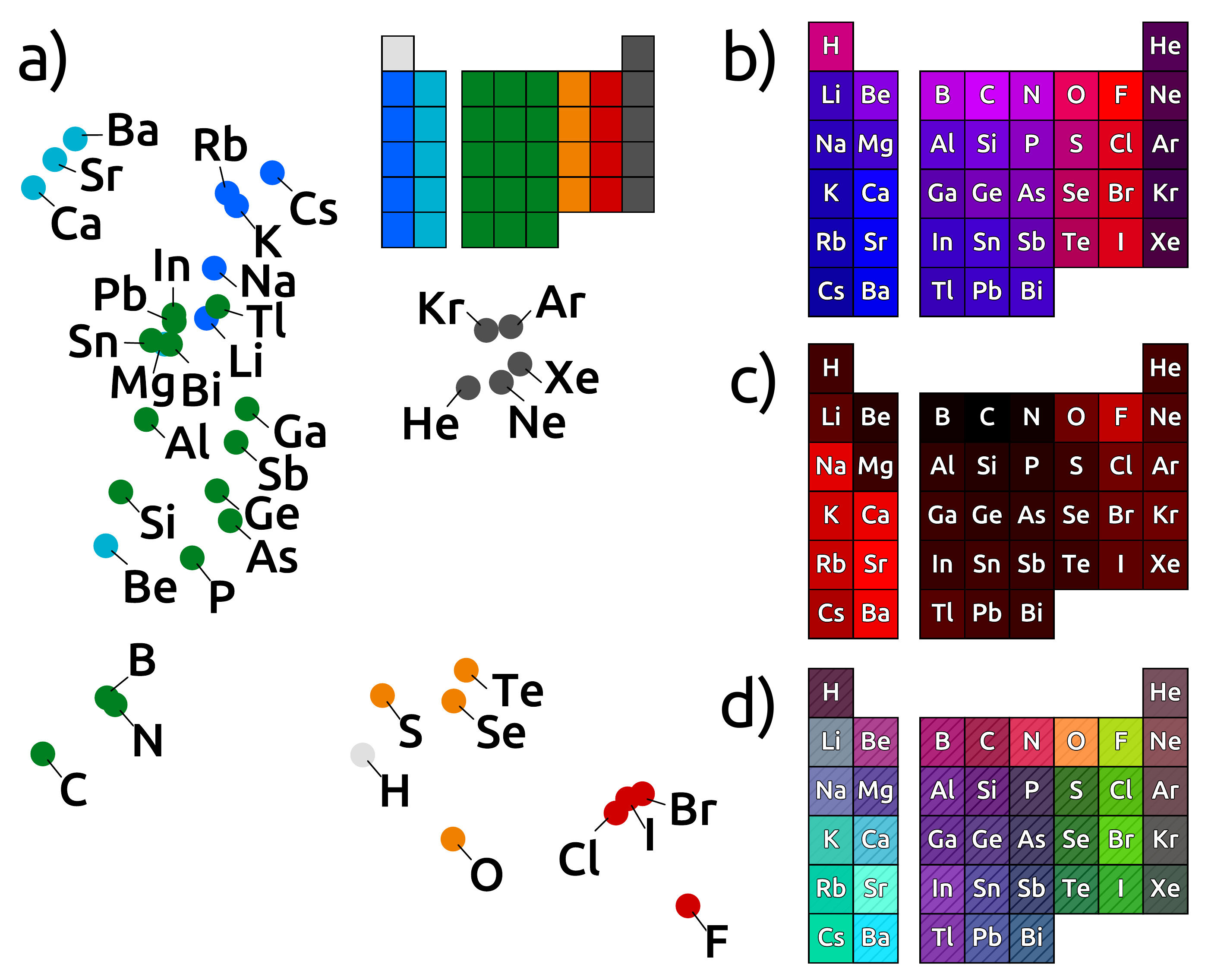}
\end{center}
\caption{Data-driven representations of the chemical space. (a) A 2D map of the elements contained in the elpasolite data set, with the coordinates corresponding to $u_{1\alpha}$ and $u_{2\alpha}$, for the case $d_J=2$. Points are colored according to the group. (b) A periodic table colored according to the coordinates in the 2D chemical space. $u_{1\alpha}$ corresponds to the red channel and $u_{2\alpha}$ to the blue channel. (c) A periodic table colored according to $u_{1\alpha}$ (red channel) for a 1D chemical space. (d)  A periodic table colored according to 4D chemical coordinates ($u_{1\alpha}$: red channel, $u_{2\alpha}$: green channel, $u_{3\alpha}$: blue channel, $u_{4\alpha}$: hatches opacity)}
\label{fig:periodictable}
\end{figure}

By increasing $d_J$ to 4 (blue line), which corresponds to a reduction in the dimensionality of the SOAP representations by 99\%, the resulting model outperforms both the reference (bright red line) and standard SOAP models. There is still, however, a reduction in the learning rate as the number of training structures increases. Again, this is likely an indication that the low dimensionality of the representation is unable to represent diversity adequately in large sets of structures (in contrast to the higher-dimensional standard SOAP representation). 

To test this idea, we combined multiple kernels in linear combination, including full-dimensionality standard SOAP kernels for $r_{c}=4, 5, 6$ and $\zeta=1, 2, 3, 4$, and the optimal alchemical kernels for $d_J = 1, 2, 4$. This multiple-kernel model (grey line) combines the optimized element correlations of the alchemical representation with the resistance to saturation of the standard SOAP representation, leading to an improvement in performance over standard SOAP and the state of the art by some 30\% on the full training set. 
It is worth noting that our regression model also outperforms by a factor of two a recently-proposed scheme to determine similarities between elements based on artificial intelligence techniques~\cite{pnas}.

The performance of the model for different levels of compression of the chemical space reflect the tradeoff between the available data and the complexity of the representation. 
Training of the extended model entails non-linear optimization of $d_J\times  n_\text{elements}$ weights, combined with KRR in a SOAP representation that contains $d_J^2$ ``element channels''. A low-dimensional model can extrapolate more reliably to combinations of elements that are not present in the train set, but may not have sufficient flexibility to maintain high learning rates when larger amounts of data are available. 
This tradeoff is evident when considering the apparent contradiction between the fact that we observed little improvement in model performance when increasing $d_{J}$ beyond four, and the fact that a multi-kernel that includes full SOAP models does improve significantly the prediction accuracy. We attribute this to the fact that the number of free parameters grows steeply with $d_{J}$, which leads to failure of cross-validation scheme to extract meaningful information from the relatively small optimization set. Conversely, the multi-kernel model provides an approach to include full element information, with only a small number of hyperparameters defining how much weight this information should be given in comparison to more coarse-grained descriptions.

\subsection{A data-driven periodic table of the elements}

The eigenvectors of the alchemical kernel $\kappa_{\alpha\alpha'}$ lend themselves naturally to be interpreted as spanning a continuous alchemical space in which the element kets $\ket{\alpha}$ are embedded. 
In other terms, they make it possible to obtain a low-dimensional representation of the elements, in which case elements that behave in a similar way with respect to the target property lie close to each other. 
Figure \ref{fig:periodictable} (a) shows the optimized distribution of the elements $u_{\alpha J}$ in the two-dimensional space spanned by $\ket{1}$ and $\ket{2}$ for $d_J=2$. Elements within different groups of the periodic table are coloured differently. 
It is immediately apparent from this colouring scheme that optimization of the alchemical kernel leads to clustering of elements that is reminiscent of their position in the periodic table. 
The correlation between the data-driven element representations and the position in the periodic table is perhaps even more apparent in Fig.~\ref{fig:periodictable} (b), in which the periodic table is color-coded according to the values of $u_{J\alpha}$.
This fascinating observation suggests that one could in principle construct a reasonable alchemical kernel using chemical intuition alone. However, there are two significant advantages to the approach presented here. First, the optimization is performed automatically on the data set under consideration. Second, the optimization can be performed just as well in a lower or higher-dimensional space (e.g. $d_J = 1$ or $d_J=4$, Fig.~\ref{fig:periodictable} (c) and (d)), where intuition based on the (two-dimensional) periodic table is likely to hinder the performance of the model. 

It should also be noted that the elpasolite data set consists of configurations that share the same structure, and span a space that is dominated by element correlations, making an optimization that ignores geometric correlations particularly effective. 
More structurally diverse data sets will imply stronger coupling between geometry and composition, making it advisable to consider more general extensions of the SOAP representations to extract comparable insight. 

\begin{figure}[tbp]
\begin{center}
    \includegraphics[width=1.0\linewidth]{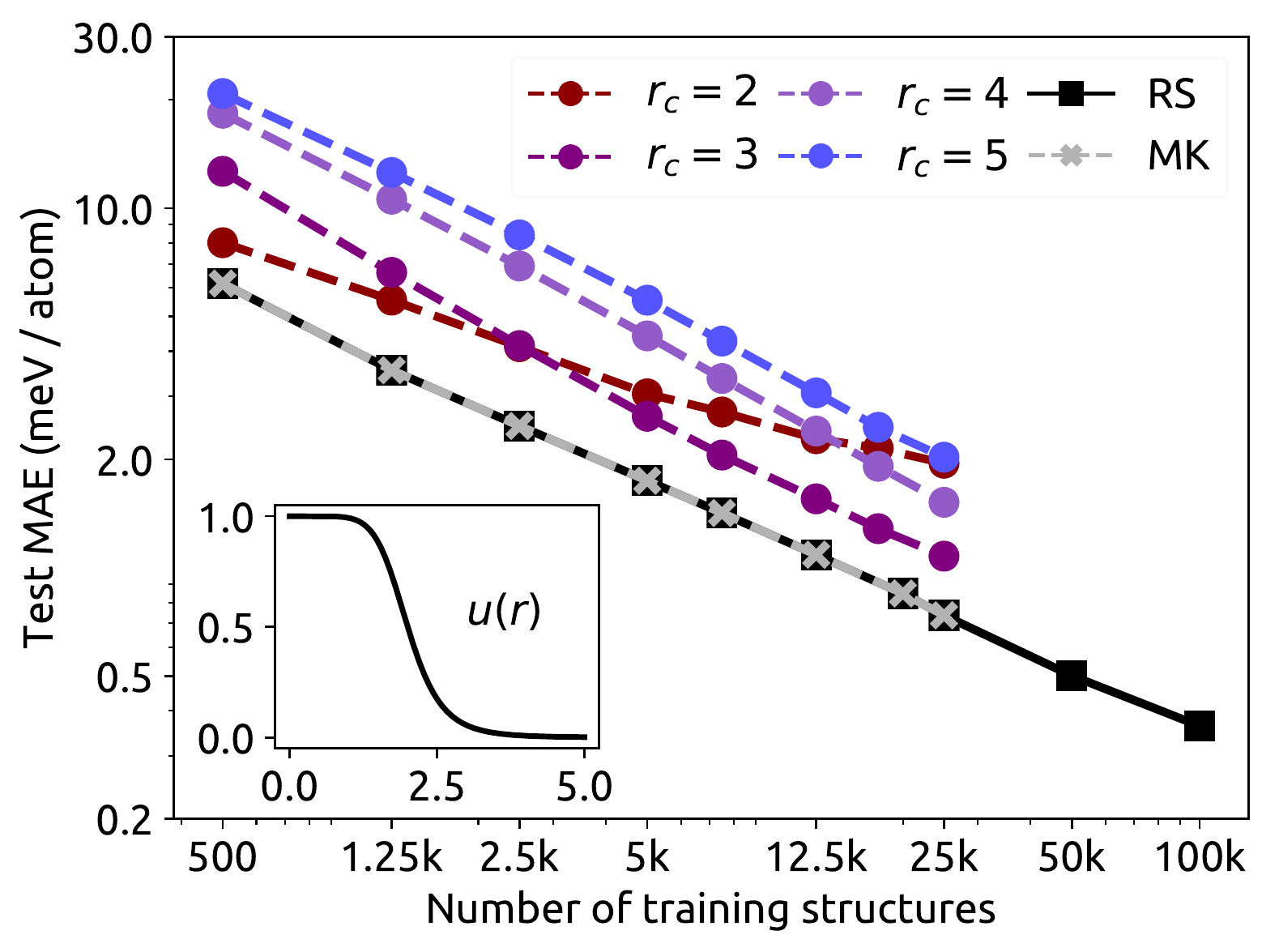}
\end{center}
\caption{Learning curves for the QM9 data set. Four of the lines show the MAE on the test set for various standard SOAP kernels ($\zeta = 2$) with different cutoff radii (dashed lines graduating from red to blue). The other lines show the MAE on the test set for the optimal radially-scaled (RS) and multiple-kernel (MK) SOAP models (black and grey lines respectively). In every model, the kernels were constructed with $\nu=2$, $n_\text{max}=12$ radial basis functions and $l_\text{max}=9$ non-degenerate spherical harmonics. The inset shows the radial-scaling function $u(r)$ from $r = 0$\AA \, to $r = 5$\AA \, with the parameters that were found to minimize the ten-fold cross validation MAE on the optimization set through a grid search, $r_{0} = 2$\AA \, and $m=7$. The multiple-kernel model combines the $r_{c}=2, 3, 4$ and RS kernels in the ratio $100,000:1:2:10,000$, and the learning curve agrees with the RS result to within graphical accuracy. Error bars are omitted because they are as small as the data point markers.
Note that errors are expressed on a per-atom basis. Error per molecule expressed in kcal/mol can be obtained approximately by multiplying the scale by 0.4147, that is computed based on the average size of a molecule in the QM9 database.
}
\label{fig:QM9-lc-rs}
\end{figure}

\subsection{Radial scaling in the QM9 data set}

Molecular databases such as the QM9 are less elementally diverse (containing only 5 elements), but contain a broad variety of structures.
It has been shown that SOAP kernels can predict with great accuracy the stability of these molecules. However, reaching the best accuracy requires a combination of kernels,  as in Eq.~\eqref{eq:multi-k}, with different cutoff radii. 
The combination of kernels with different length scales has been interpreted in terms of the need for encoding in the kernel the notion of multiple length scales in molecular interactions~\cite{Bartok2017}. 
The same argument can be applied to the optimization of a radial scaling function $u(r)$ (see Section~\ref{sub:radial}), so it should be possible to obtain similar accuracy to a multi-scale kernel by simply optimizing a suitable parameterization of such scaling. 

\begin{figure}[tbp]
\begin{center}
    \includegraphics[scale=0.50]{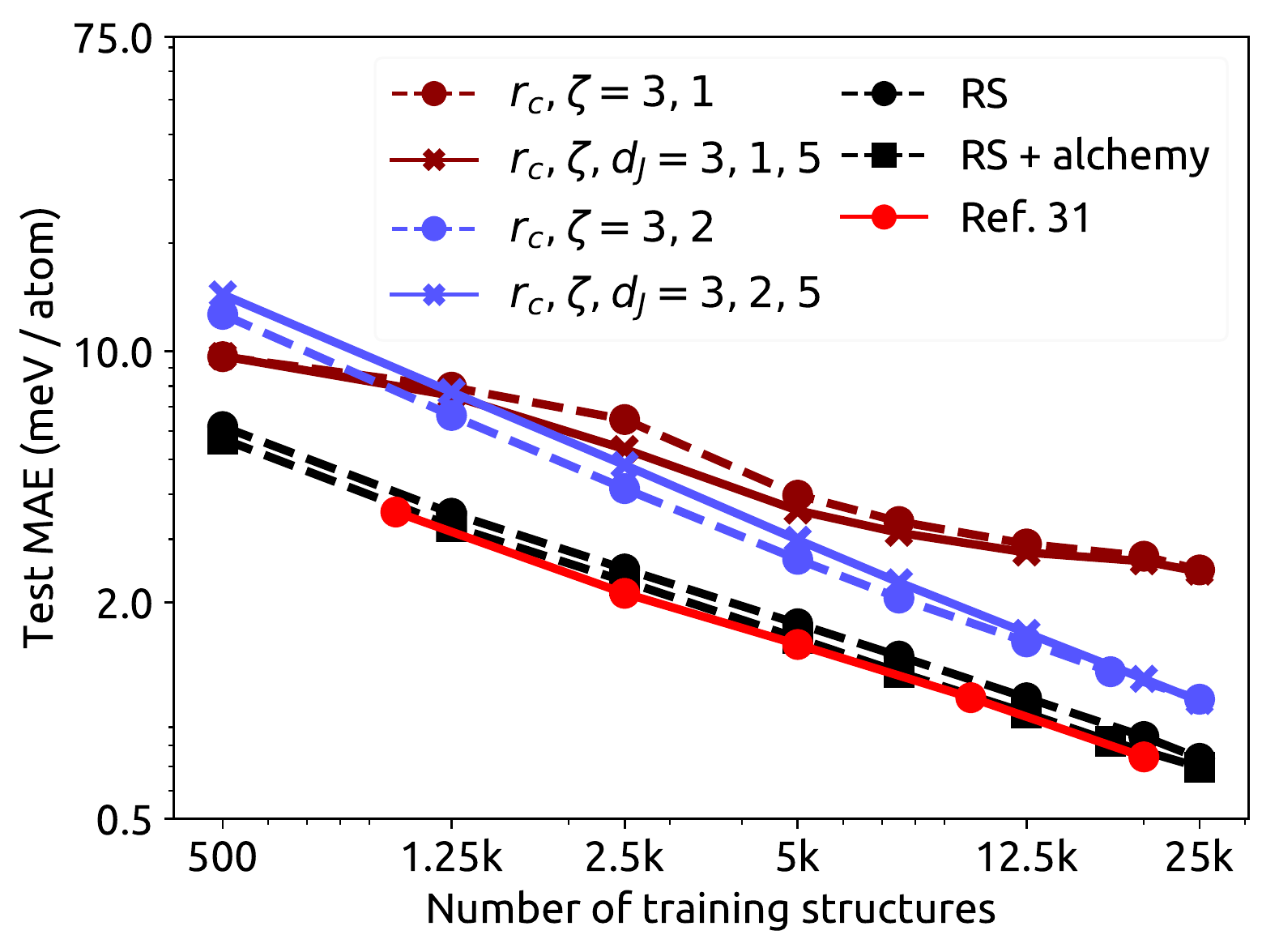}
\end{center}
\caption{Learning curves for the QM9 data set after inclusion of radially-scaled and alchemically-optimized SOAP kernels. Standard SOAP kernels with different cutoff radii are compared with the result of optimizing alchemical correlations using the scheme presented previously for the elpasolite crystal data set (blue and red lines). The learning curve of the optimized radially-scaled kernel (dashed black line with circles) is improved through inclusion of a Gaussian alchemical kernel (dashed black line with squares), which was optimized specifically for $\zeta=2$ using a grid search. The combined optimization of the radial scaling and alchemical correlations leads to a model that matches the accuracy of the state of the art curve (dashed red line), which corresponds to the representations from Ref.~\cite{Faber2018}, with the errors normalized by the average size of a molecules in the QM9 database. In every SOAP-based model, the kernels were constructed with $\nu=2$, $n_\text{max}=12$ radial basis functions and $l_\text{max}=9$ non-degenerate spherical harmonics. Error bars are omitted because they are as small as the data point markers.
}
\label{fig:QM9-lc-alchemy}
\end{figure}

Following Eq.~\eqref{eq:dirac-rxj-ur-c0}, we consider a simple functional form with a long-range algebraic decay and smooth behavior at $r\rightarrow 0$, 
\begin{equation}
u(r) = \begin{cases} 
 \frac{1}{(r/r_0)^m} &\text{if c=0,}\\
 1 &\text{if m=0,} \\
 \frac{c}{c+(r/r_0)^m} &\text{else}.
\end{cases} \label{eq:radial-u}
\end{equation}
We optimized $r_0$ and $m$ using a grid search and 10-fold cross validation over an optimization set of 5,000 randomly-selected molecules with $c=1$. The full set of parameters that we tested is given in the SI.
Figure~\ref{fig:QM9-lc-rs} compares the learning curves of conventional SOAP for different cutoff radii with the best radial scaling determined on the A set. 
Radial scaling leads to a substantial (~25\%{}) improvement in the performance of the model\footnote{It is important to stress that the results we report here are about 20\% better than those in Ref.~\cite{Bartok2017}, because we removed the 3,054 structures that failed the SMILES consistency test, as is done by other papers using this data set as benchmark, including Ref.~\cite{Faber2018}.} . The learning rate does not decrease when the training is extended to larger fractions of the QM9. At the level of 100k reference configurations, the radially-scaled kernel achieves a MAE as low as 0.34 meV/atom, corresponding to 0.14 kcal/mol. 
When considering state-of-the-art results achieved in the past year using more generally-applicable representations, our optimized model achieves an improvement which is between 25 and 60\%{}. Multi-kernel SOAP~\cite{Barker2017} yields 0.18 kcal/mol MAE, and two different neural network models reach 0.26~\cite{lubb+18jcp} or 0.32~\cite{schu+18jcp} kcal/mol MAE. 
We also attempted to build a multi-kernel model including both conventional SOAP kernels and the best radially-scaled kernels. The improvement we could achieve is marginal, which reinforces the notion that an optimal radial scaling of the representation is essentially equivalent to an optimized combination of representations with different scales.

Although the QM9 data set exhibits a low degree of composition diversity, one can attempt to further improve the performance of the model by introducing correlations between chemical species. 
In this case it is necessary to use a $\zeta=2$ exponent to incorporate many-body interactions in the regression, which makes the application of the primal-based optimization scheme we used for elpasolites impractical\footnote{Note that the $u_{J\alpha}$ optimized for the $\zeta=1$ representations lead to a degradation of the accuracy when used for the $\zeta=2$ case.}. 
For this reason, and inspired by previous results based on a heuristic determination of $\kappa_{\alpha\beta}$ based on the Pauling electronegativity of the atoms~\cite{Bartok2017}, we just used Eq.~\eqref{eq:kappa-er} and performed a grid search to find the optimal values of $\sigma_\epsilon$ and $\sigma_r$ (see the SI for more details). 
Fig.~\ref{fig:QM9-lc-alchemy} shows that this simple ansatz improves by a further 10\%{} the performance of a SOAP-based KRR model, and also combines with the optimized radial scaling to yield a model which is essentially equivalent in performance to the optimized representations of Ref.~\cite{Faber2017}. 
The success of the rather primitive form of this feature optimization protocol suggests that a more general strategy in which structural and chemical correlations are tuned simultaneously could improve even further beyond the state of the art.

\section{Conclusions}

Thanks to their mathematically sound, unbiased constructions, SOAP representations are particularly well-suited to be extended, 
incorporating information on correlations between structure, composition and properties. 
We have given two examples of such extensions, representing the behavior of different chemical species as low-dimensional vectors, and modulating the information content of the representations with a radial scaling function.
These optimizations improve significantly the performance of SOAP representations, matching or surpassing the state of the art on two very different data sets -- a chemically diverse set of quaternary solid compounds, and a collection of small organic molecules.
The framework we use to simplify the description of atomic species can reduce dramatically the complexity and computational costs of machine-learning models for multi-component systems, and could also be applied to coarse-grained models, in which beads correspond to functional groups, and a reduced-dimensionality description could identify features such as polarity or hydrophobicity. 

The exercise of optimizing SOAP representations does not only lead to more effective machine learning of molecular and materials stability. 
As we have demonstrated by re-discovering the periodic table of the elements, and extending it to one and four dimensions, it also  makes it possible to extract useful insights from the inspection of the optimal combinations of features.
When it comes to the applications of machine learning to chemistry, physics and materials science, accuracy and understanding go hand in hand.

\section*{Acknowledgements}

The Authors would like to thank G\'abor Cs\'anyi  for insightful discussion and comments on an early version of the manuscript. MC and MJW
were supported by the European Research Council under the European Union's Horizon 2020 research and innovation programme (grant agreement no. 677013-HBMAP). FM was supported by the the NCCR MARVEL, funded by the Swiss National Science Foundation.

\bibliographystyle{bibliography/aip}

\begin{thebibliography}{10}

\bibitem{behl-parr07prl}
J.~Behler and M.~Parrinello,
\newblock Phys. Rev. Lett. {\bf 98}, 146401 (2007).

\bibitem{bart+10prl}
A.~P. Bart{\'{o}}k, M.~C. Payne, R.~Kondor, and G.~Cs{\'{a}}nyi,
\newblock Phys. Rev. Lett. {\bf 104}, 136403 (2010).

\bibitem{rupp+12prl}
M.~Rupp, A.~Tkatchenko, K.-R. M{\"{u}}ller, and O.~A. von Lilienfeld,
\newblock Phys. Rev. Lett. {\bf 108}, 058301 (2012).

\bibitem{bart+13prb}
A.~P. Bart{\'{o}}k, R.~Kondor, and G.~Cs{\'{a}}nyi,
\newblock Phys. Rev. B {\bf 87}, 184115 (2013).

\bibitem{glie+17prb}
A.~Glielmo, P.~Sollich, and A.~{De Vita},
\newblock Phys. Rev. B {\bf 95}, 214302 (2017).

\bibitem{Grisafi2018}
A.~Grisafi, D.~M. Wilkins, G.~Cs{\'{a}}nyi, and M.~Ceriotti,
\newblock Phys. Rev. Lett. {\bf 120}, 036002 (2018).

\bibitem{Glielmo2018}
A.~Glielmo, C.~Zeni, and A.~De~Vita,
\newblock Phys. Rev. B {\bf 97}, 184307 (2018).

\bibitem{VonLilienfeld2018}
O.~A. von Lilienfeld,
\newblock Angew. Chemie Int. Ed. {\bf 57}, 4164 (2018).

\bibitem{bart+13prb2}
A.~P. Bart{\'{o}}k, M.~J. Gillan, F.~R. Manby, and G.~Cs{\'{a}}nyi,
\newblock Phys. Rev. B {\bf 88}, 054104 (2013).

\bibitem{Deringer2017}
V.~L. Deringer and G.~Cs{\'{a}}nyi,
\newblock Phys. Rev. B {\bf 95}, 094203 (2017).

\bibitem{Dragoni2018}
D.~Dragoni, T.~D. Daff, G.~Csanyi, and N.~Marzari,
\newblock Physical Review Materials {\bf 2}, 013808 (2017).

\bibitem{Bartok2017}
A.~P. Bart{\'{o}}k et~al.,
\newblock Sci. Adv. {\bf 3}, e1701816 (2017).

\bibitem{De2016}
S.~De, A.~P. Bart{\'{o}}k, G.~Cs{\'{a}}nyi, and M.~Ceriotti,
\newblock Phys. Chem. Chem. Phys. {\bf 18}, 13754 (2016).

\bibitem{de+16jci}
S.~De, F.~Musil, T.~Ingram, C.~Baldauf, and M.~Ceriotti,
\newblock Journal of Cheminformatics {\bf 9}, 6 (2017).

\bibitem{musi+18cs}
F.~Musil et~al.,
\newblock Chemical Science {\bf 9}, 1289 (2018).

\bibitem{VonLilienfeld2013}
O.~A. von Lilienfeld,
\newblock Int. J. Quantum Chem. {\bf 113}, 1676 (2013).

\bibitem{fabe+17jctc}
F.~A. Faber et~al.,
\newblock J. Chem. Theory Comput. {\bf 13}, 5255 (2017).

\bibitem{bhol+07nimpr}
a.~Bholoa, S.~D. Kenny, and R.~Smith,
\newblock Nuclear Instruments and Methods in Physics Research Section B: Beam
  Interactions with Materials and Atoms {\bf 255}, 1 (2007).

\bibitem{behl11jcp}
J.~Behler,
\newblock Journal of Chemical Physics {\bf 134}, 074106 (2011).

\bibitem{Chmiela2017}
S.~Chmiela et~al.,
\newblock Science Advances {\bf 3}, e1603015 (2017).

\bibitem{smit+17cs}
J.~S. Smith, O.~Isayev, and A.~E. Roitberg,
\newblock Chemical Science {\bf 8}, 3192 (2017).

\bibitem{Zhang2018}
L.~Zhang, J.~Han, H.~Wang, R.~Car, and E.~Weinan,
\newblock Physical Review Letters {\bf 120}, 143001 (2018).

\bibitem{nguy+18jcp}
T.~T. Nguyen et~al.,
\newblock J. Chem. Phys. {\bf 148}, 241725 (2018).

\bibitem{Qu2018}
C.~Qu, Q.~Yu, B.~L. {Van Hoozen}, J.~M. Bowman, and R.~A.
  Vargas-Hern{\'{a}}ndez,
\newblock Journal of Chemical Theory and Computation {\bf 14}, 3381 (2018).

\bibitem{rasm06book}
C.~E. Rasmussen,
\newblock {\em {Gaussian processes for machine learning}},
\newblock MIT Press, 2006.

\bibitem{Bishop2016}
C.~M. Bishop,
\newblock {\em {Pattern Recognition and Machine Learning}},
\newblock Springer, 2016.

\bibitem{Cuturi2010}
M.~Cuturi,
\newblock {Positive Definite Kernels in Machine Learning}, 2010.

\bibitem{arxiv_p2}
M.~J. Willatt, F.~Musil, and M.~Ceriotti,
\newblock Arxiv  (2018).

\bibitem{Bartok2013}
A.~P. Bart{\'{o}}k, R.~Kondor, and G.~Cs{\'{a}}nyi,
\newblock Phys. Rev. B {\bf 87}, 184115 (2013).

\bibitem{paruzzo2018chemical}
F.~M. Paruzzo et~al.,
\newblock arXiv preprint arXiv:1805.11541  (2018).

\bibitem{Faber2018}
F.~A. Faber, A.~S. Christensen, B.~Huang, and O.~A. {Von Lilienfeld},
\newblock Journal of Chemical Physics {\bf 148}, 241717 (2018).

\bibitem{Huang2016}
B.~Huang and O.~A. von Lilienfeld,
\newblock J. Chem. Phys. {\bf 145}, 161102 (2016).

\bibitem{QUIP}
G.~Csanyi, J.~Kermode, and N.~Bernstein,
\newblock {QUIP and quippy documentation}.

\bibitem{nong+17prb}
N.~Artrith, A.~Urban, and G.~Ceder,
\newblock Phys. Rev. B {\bf 96}, 014112 (2017).

\bibitem{gast+18jcp}
M.~Gastegger, L.~Schwiedrzik, M.~Bittermann, F.~Berzsenyi, and P.~Marquetand,
\newblock J. Chem. Phys. {\bf 148}, 241709 (2018).

\bibitem{Huo2017}
H.~Huo and M.~Rupp,
\newblock (2017).

\bibitem{goodfellow2016deep}
I.~Goodfellow, Y.~Bengio, A.~Courville, and Y.~Bengio,
\newblock {\em Deep learning}, volume~1,
\newblock MIT press Cambridge, 2016.

\bibitem{Ramakrishnan2014}
R.~Ramakrishnan, P.~O. Dral, M.~Rupp, and O.~A. von Lilienfeld,
\newblock Sci. Data {\bf 1}, 1 (2014).

\bibitem{Ramakrishnan2015a}
R.~Ramakrishnan, P.~O. Dral, M.~Rupp, and O.~A. von Lilienfeld,
\newblock J. Chem. Theory Comput. {\bf 11}, 2087 (2015).

\bibitem{fabe+18jcp}
F.~A. Faber, A.~S. Christensen, B.~Huang, and O.~A. von Lilienfeld,
\newblock J. Chem. Phys. {\bf 148}, 241717 (2018).

\bibitem{schu+18jcp}
K.~T. Sch{\"{u}}tt, H.~E. Sauceda, P.-J. Kindermans, A.~Tkatchenko, and K.-R.
  M{\"{u}}ller,
\newblock J. Chem. Phys. {\bf 148}, 241722 (2018).

\bibitem{lubb+18jcp}
N.~Lubbers, J.~S. Smith, and K.~Barros,
\newblock J. Chem. Phys. {\bf 148}, 241715 (2018).

\bibitem{Faber2016}
F.~A. Faber, A.~Lindmaa, O.~A. von Lilienfeld, and R.~Armiento,
\newblock Phys. Rev. Lett. {\bf 117}, 135502 (2016).

\bibitem{Bartok2013a}
A.~P. Bart{\'{o}}k, M.~J. Gillan, F.~R. Manby, and G.~Cs{\'{a}}nyi,
\newblock Phys. Rev. B {\bf 88}, 054104 (2013).

\bibitem{pnas}
Q.~Zhou et~al.,
\newblock Proceedings of the National Academy of Sciences {\bf 115}, E6411
  (2018).

\bibitem{Note1}
It is important to stress that the results we report here are about 20\% better
  than those in Ref.~\cite {Bartok2017}, because we removed the 3,054
  structures that failed the SMILES consistency test, as is done by other
  papers using this data set as benchmark, including Ref.~\cite {Faber2018}.

\bibitem{Barker2017}
J.~Barker, J.~Bulin, J.~Hamaekers, and S.~Mathias,
\newblock {LC-GAP: Localized coulomb descriptors for the gaussian approximation
  potential},
\newblock in {\em Scientific Computing and Algorithms in Industrial
  Simulations: Projects and Products of Fraunhofer SCAI}, pages 25--42, 2017.

\bibitem{Note2}
Note that the $u_{J\alpha }$ optimized for the $\zeta =1$ representations lead
  to a degradation of the accuracy when used for the $\zeta =2$ case.

\bibitem{Faber2017}
F.~A. Faber et~al.,
\newblock Journal of Chemical Theory and Computation {\bf 13}, 5255 (2017).

\end{thebibliography}

\end{document}